\documentclass[twocolumn,a4paper,showpacs]{revtex4}
\usepackage{graphicx}
\usepackage{amssymb}
\newcommand{\etal}{{\it et al}.}
\begin{document}

\title{Comment on ''Measurement of Effective Temperatures in an Aging Colloidal Glass''}

\author{Pierre Jop, Artyom Petrosyan \& Sergio Ciliberto}
\affiliation{Laboratoire de Physique, CNRS UMR 5276, \'Ecole Normale Sup\'erieure de Lyon, 46 all\'ee d'Italie,
69364 Lyon cedex 7, France.}

\pacs{64.70.Pf, 05.40.-a, 05.70.Ln, 82.70.Dd}

\maketitle
In a Letter \cite{Greinert07}, Greinert \etal~reported an increase of effective temperature ($T_{eff}$) during the aging process of a Laponite suspension measured from the fluctuations of position of optically trapped beads. We think their results are not conclusive since there could be some artifacts in the experiment.

We measure the fluctuations of the position of a silica bead trapped by an optical tweezers during the aging of the Laponite.
The laser beam ($\lambda$=980 nm) is focused by a microscope objective ($\times$63) 20 $\mu$m above the cover-slip surface to create a harmonic potential well where a bead of 1 or 2 $\mu$m in diameter ($d$) is trapped.
The Laponite mass concentration is varied from 1.2 to 3\% wt for different ionic strengths.
 
Following the method used in \cite{Greinert07}, the stiffness of the optical trap is periodically switched between two different values ($k_1$ and $k_2$) every 61 seconds by changing the laser intensity. The position of the bead is recorded by a quadrant photodiode at the rate of 8192 Hz, then we compute the variance of the position over the whole signal, $\left<\delta x^2\right>=\left< x^2-\left<x\right>^2\right>$, where the brackets stand for average over the time. To avoid transients, each record is started 20 seconds after the laser switch. 
 Assuming the equipartition principle still holds in this out-of-equilibrium system, the $T_{eff}$ is computed as in \cite{Greinert07}.

\begin{figure}[h!]
	\centering
		\includegraphics[width=8.5cm]{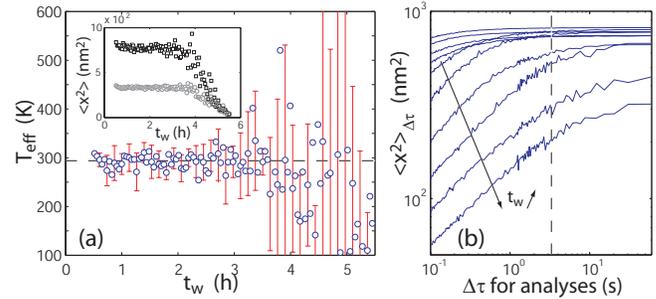}
\caption{\label{fig:errorbarssdet}a) Evolution of the effective temperature with aging time for $k_1=6.34$ pN/$\mu$m, $k_2=14.4$ pN/$\mu$m, 2.3\% wt of Laponite and 1 $\mu$m glass bead. The error bars are computed from the statistical error of $\left<\delta x_i^2\right>$ at the time $t_w$. Inset: Evolution of $\left<\delta x^2\right>$ with $t_w$ for both stiffnesses ($k_1$ $\circ$ and $k_2$ $\square$). b) Evolution of $\left<\delta x^2\right>_{\Delta \tau}$ for the low stiffness as a function of the duration $\Delta \tau$ of the samples (see text) for different aging times : from top to bottom $t_w=30$, $57$, $83$, $110$, $137$, $163$, $190$, $217$, $243$ and $270$ minutes.}
\end{figure}

 Our results are shown on the Fig.~\ref{fig:errorbarssdet}a. We find that $T_{eff}$ is constant at the beginning and is very close to $T_{bath}=294$ K, 
 then when the jamming occurs, it becomes noisy without any clear increase with $t_w$, contrary to Ref. \cite{Greinert07}.

We now make three remarks. First, we point out that the uncertainty of their results are underestimated. The error bars in Fig. \ref{fig:errorbarssdet}(a) are here evaluated from the standard deviation of the variance using Eq.~2 in Ref \cite{Greinert07} at the time $t_w$. Although they are small for short time $t_w$, ($\Delta  T_{eff}/T_{eff}\leq 10\%$), they increase for large $t_w$. This is a consequence of the increase of variabilities of $\left<\delta x_i^2\right>$ as the colloidal glass forms. This point is not discussed in in Ref. \cite{Greinert07} and we think that the measurement errors are of the same order or larger than the observed effect.  
The results depend on the length of the analysing time window and the use of the principle of energy equipartition becomes questionable for the following reasons.

First, these analysing windows cannot be made too large because the viscoelastic properties of Laponite evolve as a function of time. 
 Second, the corner frequency of the global trap (optical trap and gel), the ratio of the trap stiffness to viscosity, decreases continuously mainly because of the increase of viscosity. At the end of the experiment, the power spectrum density of the displacement of the bead shows that the corner frequency is lower than $0.1$ Hz.  
We thus observe long lived fluctuations, which could not be taken into account with short measuring times. This problem is shown on Fig.~\ref{fig:errorbarssdet}b. We split our data into equal time duration $\Delta \tau$, compute the variance and average the results of all samples. The dotted line represents the duration 3.3 s chosen in \cite{Greinert07}.
At the beginning of the experiment, the variance of the displacement is constant for any reasonable durations of measurement. However, we clearly see that this method produces an underestimate of $\left<\Delta x^2\right>$ for long aging times, specially when the viscoelasticity of the gel becomes important. Long lived fluctuations are then ignored.

  Finally, a remark has to be made concerning the mean position of the bead during aging. When the stiffness of the gel becomes comparable to the optical one, the bead starts to move away from the center the optical trap. We observe a drift of the bead position at long time, which could lead to the escape of the bead. Moreover we have performed simultaneous measurements with a multiple trap using a fast camera showing that at very long $t_w$ the mean trajectories of beads separated by 7 $\mu$m are almost identical. This proves that one must pay attention when interpreting such measurements, specially on the duration of measurements.
We also have seen that the way the sample is sealed can accelerate the formation of the gel and the drift of the bead by changing the chemical properties in the small sample. We have used different types of cell, Laponite concentrations, bead sizes, stiffnesses of the optical trap. In each case we do not find any increase of the effective temperature.

In conclusion, our results show no increase of $T_{eff}$ in Laponite and are in agreement with those of Jabbari-Farouji \etal \cite{Jabbari-Farouji07}, who measured fluctuations and responses of the bead displacement in Laponite over a wide range of frequency and found that $T_{eff}$ is equal to the bath temperature.


\end{document}